# A Cinemática do Movimento Circular
*(The Kinematics of Circular Motion)*

L. A. N. de Paula[1]
Universidade Federal de Itajubá - UNIFEI[2]
23 de Maio de 2008

**Resumo**

Neste artigo introduzimos uma proposta para a cinemática dos corpos em movimento circular uniforme. Este modelo poderia contribuir para a explicação de dois dos principais problemas da cosmologia atual: matéria escura e energia escura. Utilizamos uma das propriedades físicas do espaço – as ondas gravitacionais – para definir um sistema de referência em repouso no espaço e determinar um mecanismo para a sincronização de relógios. Dessa forma derivamos, através de um Postulado, as conhecidas transformações de Lorentz e uma nova velocidade para os corpos em movimento de rotação uniforme. Esta velocidade de rotação contribuiria na elucidação dos dois problemas cosmológicos. Essa contribuição viria de efeitos devidos a um único fenômeno: a velocidade de rotação dos corpos em relação ao sistema de referência em repouso no espaço. E por fim, interpretamos fisicamente o Postulado - pilar deste modelo - que permite, inclusive, uma comparação com o Princípio de Relatividade e também novas previsões teóricas.

**Palavras-chave:** Matéria Escura, Energia Escura e Movimento Circular Uniforme.

**Abstract**

In this paper we introduce a proposal for the kinematics of bodies in uniform circular motion. This model could contribute for the explanation of the two main problems of contemporary cosmology: dark matter and dark energy. We use one of the physical properties of space – the gravitational waves – to define a reference frame at rest with respect to the space and to determine a mechanism for the synchronization of clocks. So we derive through a Postulate the so-called transformations of Lorentz and a new velocity of the bodies in uniform circular motion. This rotational velocity could contribute to clarify both cosmological problems. This contribution would come from effects due to a single phenomenon: the rotational velocity of bodies with respect to the reference frame at rest on the space. And finally, we interpret physically the Postulate – the major point in this model – which allows, inclusive, a comparison with the Relativity Principle and also new theoretical previews.

**Keywords:** Dark Matter, Dark Energy and Uniform Circular Motion.

---

[1] e-mail: leandroifgw@yahoo.com.br
[2] Instituto de Ciências, Av. BPS 1303 Pinheirinho, 37500-903 Itajubá, MG, Brasil.

## 1. Introdução

No decorrer do século passado, foram introduzidas muitas novidades na ciência. Particularmente, na física, assistimos ao advento da ascensão da cosmologia. À medida que o tempo avança tem sido possível cada vez mais sondar partes na natureza que antes eram impensadas devido aos parcos recursos tecnológicos. Como todo o ramo da ciência, nos seus primórdios, a cosmologia atual tem encontrado muitas dificuldades na explicação de dados astrofísicos. Alguns dos principais responsáveis por essa deficiência seriam os recursos oferecidos pelas teorias vigentes. Toda teoria vigente precisa, cedo ou tarde, ser ampliada à medida que ampliamos nossos conhecimentos sobre o universo.

As cinemáticas e dinâmicas, que atualmente tomamos como referência para a explicação do cosmo, foram embasadas em uma época em que pouco se sabia a respeito dos fenômenos extragalácticos. Seja, por exemplo, o movimento de rotação de corpos no halo de uma galáxia, situados a uma distância da ordem de dez mil anos-luz de seu centro. A cinemática newtoniana prevê que a velocidade de rotação destes corpos diminui com o aumento da distância ao centro. Contudo, para distâncias dessa magnitude esta lei falha. Da mesma forma a cinemática einsteniana não enquadra estes tipos de movimento. Outros fenômenos, como o aumento do desvio espectral para o vermelho com o aumento da distância entre as galáxias, também não têm sido satisfatoriamente explicados. Para preencher estas lacunas muitas propostas podem ser feitas.

Em particular, propomos que a cinemática do movimento circular deve ser modificada sensivelmente nas escalas extragalácticas. Na seção seguinte, fazemos uma breve menção sobre o desenvolvimento da concepção de um meio físico que permeia todo o universo, que atualmente denominamos de 'espaço dotado de propriedades físicas'. Na seção 3, utilizamos uma dessas propriedades, que são as ondas gravitacionais, como um mecanismo para a identificação de um sistema de referência em repouso no espaço. Nas seções 4 e 5, esclarecemos o que entendemos por sistema de referência e sincronização de relógios. E ao identificarmos o sistema de referência em repouso no espaço, por meio de um Postulado, derivamos as conhecidas transformações de Lorentz. Elas nos serão úteis para derivarmos, na seção 6, uma nova velocidade para os corpos em movimento circular uniforme em relação a um sistema de referência em repouso no espaço. E nas últimas seções, utilizamos esta velocidade de rotação para entendermos fenômenos como aqueles denominados matéria escura e energia escura.

## 2. Espaço e Ondas Gravitacionais

Até fins do século XIX, o conceito de éter foi sendo gradativamente elaborado por renomados e brilhantes cientistas tais como Oersted, Faraday, Lorentz, Poincaré e principalmente James C. Maxwell [1]. Para este o éter era uma substância dotada de propriedades físicas que preenchia todo o espaço vazio, sendo o carregador das ondas eletromagnéticas. Em 1887, Heinrich Hertz produziu experimentalmente estas ondas e, a partir disso, praticamente todos os físicos passaram a acreditar que o éter realmente existia. Michelson e Morley elaboraram um experimento extremamente perspicaz para detectar o "vento etéreo": ou o movimento do éter em relação à Terra [2]. Porém, o movimento relativo esperado não foi detectado. Lorentz, apoiado no resultado deste experimento e na crença da existência dessa substância, construiu uma hipótese sobre a contração dos corpos em movimento relativo ao éter.

Em 1905 Einstein publicou sua teoria que viria mais tarde persuadir a maior parte dos físicos a desacreditarem na existência dessa substância. Esta teoria, apoiada no princípio de relatividade levantada outrora por Poincaré, demonstrava que era possível derivar de modo mais simples os resultados até então já encontrados através do conceito de éter. O éter passou a cair em descrédito, sendo que o próprio Einstein passou a ignorá-lo.

Porém, a partir de 1920, após a formulação da teoria da relatividade geral, um novo conceito estava sendo formulado [3]. Segundo Einstein, o éter era o próprio espaço dotado de propriedades físicas. Este novo conceito iria acompanhá-lo até o final de sua vida em 1955. Em 1954, um ano antes de sua morte, ele disse: *"Espaço físico e éter são somente termos diferentes para a mesma coisa; campos são estados físicos do espaço,"* [4].

Mas por que os físicos contemporâneos criticam, de forma até mesmo preconceituosa, a sua existência? Criticam com motivo, uma vez que uma versão mais elaborada do éter passou a ser formulada: o espaço dotado de propriedades físicas. O éter não deve ser pensado como uma substância estendida no espaço, mas sim ele é o próprio espaço. Quando os físicos modernos se referem ao espaço com suas propriedades físicas estão, até mesmo sem saber, dando apoio à construção de um conceito que vem sendo elaborado desde séculos passados. Realmente não deveríamos mais falar em éter, como sendo algo estendido no espaço, mas sim no espaço dotado de propriedades físicas.

De acordo com a descrição da teoria da relatividade geral, o espaço pode sofrer deformações e, além disso, pode transportar energia. Uma propriedade dinâmica do espaço que tem sido explorada recentemente são as ondas gravitacionais. Pois assim como cargas aceleradas produzem ondas eletromagnéticas, massas aceleradas também produzem ondas gravitacionais, que se propagam no espaço com a mesma velocidade que as ondas eletromagnéticas. Em 1993, Hulse e Taylor receberam o prêmio Nobel de Física. Eles conseguiram detectar, pela primeira vez, um indício indireto da existência das ondas gravitacionais. Em 1974, eles descobriram a existência de um pulsar binário (*estrelas de nêutrons*) denominado PSR 1913+16 [5]. Após coletarem os dados do período de rotação desse pulsar ao longo de vários anos, verificaram uma taxa de decaimento nesse período. Relacionando a massa e o período do pulsar puderam calcular um decaimento na massa desse sistema binário. Incrivelmente, esta perda de energia massiva correspondia precisamente à perda de energia através de uma dissipação por ondas gravitacionais, prevista pelo modelo da teoria da relatividade geral de Einstein [6]. Dessa forma, as ondas gravitacionais foram detectadas somente de modo indireto. Recentemente, alguns detectores já estão até mesmo em operação para sua detecção direta.

A confirmação das ondas gravitacionais seria extremamente importante para as aplicações na área das radiações. E seria de especial interesse neste trabalho. Pois utilizaremos estas ondas que se propagam no espaço com velocidade igual a da luz para, por exemplo, identificarmos um sistema de referência em repouso no espaço. Também utilizaremos estas ondas ao adotarmos um procedimento para a sincronização de relógios.

3. **Identificação do Sistema de Referência em Repouso no Espaço**

Muitos cientistas no passado questionavam qual o corpo ou sistema de referência que estava em repouso no éter. Da mesma forma, devemos indagar novamente qual o corpo ou sistema de referência que está em repouso no espaço. Antes de tudo é preciso esclarecer o que entendemos por sistema de referência e repouso.

Um sistema de referência é aquele através do qual qualquer corpo material pode ser localizado no espaço de forma precisa. Visualize, por exemplo, uma colina sendo que nas suas regiões mais baixas foram construídas duas rodovias retas e que se cruzam em um determinado trecho perpendicularmente. E dentre todos os postes de iluminação que foram implantados nestas rodovias, existe um que é comum a elas e está localizado justamente no cruzamento. Suponhamos que se queira fazer um mapa para a localização de um objeto enterrado na colina. Uma possível maneira de localizar este objeto no mapa seria contar o número de passos do objeto até uma das rodovias, ao caminhar perpendicularmente em relação a ela. E em seguida, contar o número de passos até o poste ao andar paralelo a esta rodovia. E por fim, mede-se o número de palmos no poste, do solo até a altura do objeto. O caminho inverso do poste até o objeto poderia ser realizado por qualquer pessoa que possuísse o mapa. Esta localização é possível porque se construiu um sistema de referência: as duas rodovias, o poste, o número de passos e palmos. Poderíamos aplicar este método, na imaginação, para localizar qualquer objeto em algum lugar prolongando as extremidades das rodovias e do poste até o infinito. Para fins matemáticos, podemos associar a esta noção de sistema de referência um sistema de coordenadas cartesiano.

A noção intuitiva de repouso relativo é clara. Quando não há movimento relativo entre dois corpos dizemos que eles estão em repouso entre si. Entretanto, vamos definir uma outra espécie de repouso. Para isto precisaremos de um aparato experimental similar ao interferômetro de Michelson-Morley (IM).

Sabemos que o interferômetro de Michelson-Morley foi originalmente projetado para detectar o movimento relativo entre o éter e a Terra. É constituído por dois braços longos e perpendiculares que servem de suporte para o caminho de um feixe de luz. Através de divisões, reflexões e diferenças de caminho deste feixe, podem ser observadas franjas de interferência em um anteparo. O desvio destas franjas para um determinado valor poderia ser observado caso houvesse movimento entre a Terra e o éter. Pois se supunha que a luz propagava-se neste meio com uma velocidade definida e constante. Durante o século XX, outros interferômetros foram projetados, porém, após uma análise estatística dos dados bem acurada, todos estes experimentos parecem não ter produzido o resultado esperado. Como sabemos que as ondas gravitacionais se propagam no espaço com uma velocidade definida e constante igual a da luz, um aparato experimental também poderia ser projetado para detectar o movimento relativo entre a Terra e o espaço. Podemos, por exemplo, denominar este aparato de interferômetro adaptado às ondas gravitacionais (IG). Conseguimos, agora, estabelecer o conceito de uma outra espécie de repouso.

Consideremos um sistema de referência qualquer, um IG e um observador em repouso relativo na origem deste sistema. Suponha que após a coleta dos dados deste IG, este observador constata um resultado nulo $(IG \equiv 0)$. Isso significa que ele mediu uma velocidade igual a da luz para as ondas gravitacionais. Portanto, este sistema de referência está localmente em repouso no espaço. Ou seja, pelo menos em uma região indeterminada nas vizinhanças do IG o espaço não possui movimento neste sistema de referência. Denominaremos este tipo de repouso de repouso local.

Podemos igualmente identificar o repouso global, ou apenas, repouso. É necessário colocar IG's em repouso relativo no sistema de referência em todas as posições possíveis deste sistema. Se um observador situado na origem ao tratar os dados dos IG's constata um resultado nulo ($IG's \equiv 0$) para todos, então este sistema de referência está em repouso global. Chamaremos um sistema de referência em repouso global de Sistema de Referência em Repouso no Espaço (SR). Evidentemente, a implementação prática desta definição é inviável. Por isso, introduziremos um postulado mais adiante para a identificação de um SR.

Sabendo onde está um SR por meio de um postulado, podemos afirmar que um IG em repouso relativo neste SR produzirá um resultado nulo. E isto qualquer que seja a posição do IG neste sistema. Contrariamente, se um dos $IG's \equiv 0$ entrar em movimento neste sistema, seu resultado não será nulo; será dependente dessa sua quantidade de movimento. Contudo, para realizar qualquer medida dessa natureza em um sistema de referência, é necessário, antes, definir padrões para as medidas de comprimento e tempo.

### 4. **Medidas de Comprimento e Tempo**

Para que possamos fazer medidas da posição de um corpo material em todo instante de tempo é preciso definir e estabelecer padrões de medida de comprimento e tempo em um sistema de referência.

Podemos substituir os passos e palmos que utilizamos para marcar a posição do objeto na colina por uma régua graduada, cuja unidade de medida seja o metro. Ao invés de contarmos o número de passos e palmos, colocamos uma extremidade da régua onde estava a outra sucessivamente, do objeto até a rodovia, desta até o poste e daí até a altura do objeto. Teremos ao final deste processo três comprimentos ou números em metros que darão a localização unívoca do objeto na colina. É claro que, matematicamente, associamos a este sistema de referência, com este procedimento de medida de comprimento, um sistema de coordenadas cartesianas cujas unidades são dadas em metro.

Por outro lado, para estabelecermos um padrão de medida de tempo em um sistema de referência inevitavelmente precisamos antes esclarecer o conceito de simultaneidade. Devemos entender de que maneira marcamos o tempo para os eventos ou acontecimentos que ocorrem no espaço. Quando dizemos que algo aconteceu às sete horas - a parada de um ônibus próximo ao poste no cruzamento das rodovias às sete horas, por exemplo - estamos dizendo que a parada do ônibus coincide com a posição dos ponteiros do relógio, em sete horas, de um observador também nas vizinhanças do poste: dizemos que são eventos simultâneos. Porém, este procedimento para a medida de tempo só pode ser realizado para eventos que ocorrem nas vizinhanças do observador com o seu relógio de ponteiros.

Para observadores que estejam muito distantes do evento saibam a que horas o evento ocorreu no relógio do observador próximo ao evento é preciso que estes relógios estejam sincronizados através de um procedimento de sincronização. Um exemplo de procedimento de sincronização são os fusos horários. Alguém que esteja assistindo no Brasil uma notícia ao vivo no Japão pela televisão, sabe que estes acontecimentos ocorrem lá com doze horas de diferença devido aos fusos.

Suponhamos dois observadores A e B, cada um com o seu relógio nas suas proximidades. Vamos supor que A esteja sobre a superfície da Terra e B em alguma outra galáxia longínqua, que possa ser considerada em repouso em relação à nossa galáxia em escala universal. Pois a variação das posições relativas das galáxias em um certo período de tempo relativamente curto é insignificante em vista das grandes distâncias que as separam. Eventos que ocorrem nessa galáxia, como pulsos gravitacionais de pulsares, podem ser medidos com um relógio próximo do observador B como eventos simultâneos aos ponteiros do seu relógio. Estes pulsos gravitacionais viajam no espaço à velocidade da luz e podem ser detectados aqui na Terra. Devemos sincronizar o relógio de A com o relógio de B através destas ondas, para que A possa saber o momento mais exato que o pulso ocorreu. Esta implementação de sincronização exige que o observador A tenha conhecimento de qual é a medida da velocidade dessas ondas em um referencial em repouso relativo a ele.

Neste momento, devemos fazer uma hipótese. Suponhamos que o *sistema de referência em repouso relativo ao observador é um sistema de referência em repouso no espaço (SR)*. Um observador sempre medirá uma velocidade $c$ igual a da luz para as ondas gravitacionais neste sistema de referência. Em outras palavras, o observador está em repouso no espaço. E este deve ser o único postulado que adotaremos. Contudo, devemos avisar você que não faremos nenhuma definição de observador. Pois não estamos nos referindo ao observador tal qual ele nos parece ser, mas estamos nos referindo a uma forma mais essencial de observador. Tomaremos, portanto, observador como um conceito primitivo, ou seja, é um conceito que não pode ser definido.

Assim, se o pulso ocorre no instante $t_B$ medido no relógio de B, o observador A deverá medir um tempo $t_A = l_B/c + t_B$, após as ondas gravitacionais percorrerem uma distância $l_B$ até a nossa galáxia. O momento do evento $t_B$ fica automaticamente conhecido pelo observador A e dado por $t_B = t_A - l_B/c$. Isso significa que o evento ocorre em um tempo $l_B/c$ menor do que o tempo $t_A$ medido pelo relógio do observador A. Dois eventos que ocorrem em posições distintas, $l_B$ e $l_C$ distantes de um observador A, nos instantes $t_B$ e $t_C$ medidos por relógios de observadores B e C próximos a estes eventos, respectivamente, podem não ser simultâneos, apesar de serem medidos como simultâneos pelo observador A. Para que eles de fato sejam simultâneos devemos ter $t_B = t_C$, ou $t_{AB} - t_{AC} = (l_B - l_C)/c$, onde $t_{AB}$ e $t_{AC}$ são os tempos medidos por A para os eventos B e C, respectivamente. Ou seja, para que os eventos sejam simultâneos a diferença de tempo entre eles medido por A deve ser $(l_B - l_C)/c$. Daí vem que, se $l_B \cong l_C \cong 0$ então $(l_B - l_C)/c \cong 0$ resultando em $t_{AB} \cong t_{AC}$. Portanto os eventos vizinhos podem ser definidos para terem $l_B \cong l_C \cong 0$ de forma que $t_{AB} \cong t_{AC}$. Por exemplo, se o observador A estiver a uma distância de 300 km da posição onde ocorreu o pulso, então a leitura dos ponteiros do seu relógio pode ser considerada simultânea a ocorrência do evento por erros da ordem de milisegundos. A detecção se confunde com a real ocorrência dos eventos.

Posicionando o sistema de referência de forma que o observador esteja sempre na sua origem, então a cada ponto com distância $l$ da origem do sistema de coordenadas associado podemos relacionar um relógio que marca um tempo $t = t_o - l/c$ com relação ao tempo $t_o$ medido pelo relógio do observador na origem.

Por que não usamos as ondas eletromagnéticas para a identificação do sistema de referência em repouso no espaço e para a sincronização de relógios? A resposta está no fato de que não temos uma teoria consistente onde os campos elétricos e magnéticos são vistos como propriedades físicas do espaço. Estes campos, dissociados do espaço dessa maneira, ainda pode fazer pensar na existência de um éter luminífero estendido sobre o espaço e carregador das ondas eletromagnéticas. Como já relatamos este éter luminífero pode ser o próprio espaço físico, mas vale ressaltar que ainda não há uma identificação completa.

## 5. Transformações de Coordenadas em Movimento de Translação Uniforme

Tomemos dois sistemas de referência $O$ e $O'$ dotados de IG's, réguas e relógios idênticos, em movimento de translação uniforme relativos com velocidade $v$. Pelo Postulado que adotamos, se considerarmos um observador em repouso em cada um dos sistemas de referência, então estes sistemas de referência são sistemas de referência em repouso no espaço. Assim cada observador medirá uma velocidade igual a da luz para as ondas gravitacionais em relação ao sistema de referência no qual ele está em repouso.

Associemos a cada um dos sistemas de referência $O$ e $O'$ um sistema de coordenadas cartesianas $S$ e $S'$, respectivamente, de forma que a especificação de um evento em uma posição e instante de tempo quaisquer em $O$ e $O'$ sejam identificados por $x, y, z, t$ e $x', y', z', t'$ nos sistemas de coordenadas correspondentes. Precisamos determinar as relações entre as coordenadas 'com linha' e as coordenadas 'sem linha' entre os dois sistemas. Supondo que o espaço e o tempo são homogêneos as equações que relacionam estas variáveis devem ser lineares [7]. Este argumento se justifica observando que, na ausência de forças externas, se no sistema $S'$ um corpo descreve um movimento retilíneo, no sistema $S$ este movimento deve ser visto também como um movimento retilíneo.

Adotemos um posicionamento relativo entre $S$ e $S'$ de modo que estes dois sistemas de coordenadas tenham seus eixos $x$ e $x'$ coincidentes, além de $y$ e $y'$ ($z$ e $z'$) serem paralelos em todo instante de tempo. E assumamos que no instante $t = 0$ medido em $S$, os dois sistemas são completamente coincidentes. Assim as seguintes equações são equivalentes

$$x' = 0 \quad x = vt \quad (x - vt = 0)$$
$$y' = 0 \quad y = 0$$
$$z' = 0 \quad z = 0$$

Elas descrevem a origem de $S'$ nos dois sistemas.

Devido à linearidade das equações devemos ter

$$x' = a(x - vt)$$
$$y' = by$$
$$z' = cz$$

E o Postulado que foi feito mais acima fornece as seguintes equações para as ondas gravitacionais em $S$ e $S'$

$$x^2 + y^2 + z^2 = c^2 t^2$$
$$x'^2 + y'^2 + z'^2 = c^2 t'^2$$

onde $c$ é a velocidade dessas ondas nos sistemas $S$ e $S'$.

Dessas equações segue que

$$t' = \varphi(v)\beta\left(t - vx/c^2\right)$$
$$x' = \varphi(v)\beta(x - vt)$$
$$y' = \varphi(v) y$$
$$z' = \varphi(v) z$$

onde $\beta = 1/\sqrt{1 - v^2/c^2}$.

Para determinar a função $\varphi(v)$, devemos considerar um outro sistema de coordenadas $S''$ se movendo em relação a $S'$ (e orientado como $S'$ está em relação a $S$) com velocidade $-v$. Assim $S$ e $S''$ são coincidentes e portanto

$$t'' = \varphi(v)\varphi(-v) t$$
$$x'' = \varphi(v)\varphi(-v) x$$
$$y'' = \varphi(v)\varphi(-v) y$$
$$z'' = \varphi(v)\varphi(-v) z$$

Logo $\varphi(v)\varphi(-v) = 1$. Como a relação entre $y$ e $y'$ não depende do sinal de $v$: $\varphi(v) = \varphi(-v)$. Assim $\varphi_{\pm}(v) = \pm 1$ e descartaremos a solução com sinal negativo obviamente.

E, portanto, as equações de transformação de coordenadas são

$$t' = \beta\left(t - vx/c^2\right)$$
$$x' = \beta(x - vt)$$
$$y' = y$$
$$z' = z$$
(1)

Podemos substituir, corretamente, as 'grandezas com linha' pelas 'grandezas sem linha' se ao mesmo tempo invertermos o sinal de $v$. Dessa forma teremos as coordenadas do sistema $S$ dadas em termos das coordenadas do sistema $S'$.

## 6. Sistemas de Coordenadas em Movimento Circular Uniforme

Tomemos diversos sistemas de referência (e rígidos) em repouso relativo uns aos outros. Todos no mesmo plano. Como exemplo, sejam estes referenciais fixos nas galáxias dispersas no Universo. E adotemos, como antes, que elas possam ser consideradas em repouso umas em relação às outras em vista de suas distâncias e velocidades relativas. Sejam todos estes referenciais orientados do mesmo modo, ou seja, os eixos $x, y, z$ de sistemas de coordenadas associados são paralelos uns em relação aos outros.

Consideremos apenas um destes sistemas de referência em movimento circular uniforme em relação aos demais sistemas situados nas outras galáxias. Para que observadores em repouso nos demais referenciais galácticos possam identificar e quantificar este movimento de rotação é preciso que eles meçam algum evento fixo neste referencial. Seja este referencial $O$ fixo na galáxia e movendo-se junto com ela em seu movimento circular e uniforme.

Escolhamos estes eventos para serem pulsos gravitacionais com localidade próxima à origem do sistema de coordenadas associado ao referencial $O$ em rotação. Seja um deles ocorrendo no instante $t_A$ e outro no instante posterior $t_B$, medido no relógio de um observador próximo aos pulsos gravitacionais. Um outro observador, em repouso nas proximidades da origem de qualquer um dos outros referenciais, pode medir o movimento de rotação do referencial $O$ ao identificar estes dois pulsos nas proximidades da origem do seu referencial $O'$. No instante $t'_C$ ele identifica o pulso gravitacional que parte das proximidades da origem do referencial galáctico $O$ no instante $t'_A$, radialmente para a direção do seu referencial galáctico $O'$. E no instante posterior $t'_D$ ele detecta o outro pulso gravitacional que também parte das proximidades da origem de $O$ no instante $t'_B$ radialmente para o seu referencial galáctico[3].

Assim o movimento de rotação do referencial $O$ pode ser medido como segue. Pelo procedimento de sincronização adotado aqui, o observador em $O'$ sabe que os eventos de fato ocorrem em $t'_A = t'_C - r'_C/c$ e $t'_B = t'_D - r'_D/c$; onde $r'_C$ e $r'_D$ são as distâncias dos pontos de partida dos pulsos até os pontos na proximidade da origem do referencial $O'$ onde estes pulsos são medidos nos instantes $t'_C$ e $t'_D$, respectivamente. Como podemos sempre fazer $r' = r'_C = r'_D$ por uma escolha conveniente dos pontos de partida e chegada dos pulsos, onde $r'$ é a distância das origens destes sistemas associados aos referenciais galácticos $O$ e $O'$, teremos sempre $t'_D - t'_C = t'_B - t'_A$. Portanto, o intervalo de tempo entre os dois pulsos medidos em qualquer dos outros referenciais é sempre o mesmo, pois ele é igual ao intervalo de tempo entre os instantes de partida dos pulsos. Seja também $\Delta\theta_o$ o ângulo medido no referencial $O'$ definido pelos pontos de chegada dos pulsos nos instantes $t'_C$ e $t'_D$ e a origem do referencial $O$. A velocidade de rotação do referencial $O$ pode ser definida em $O'$ para ser $\dot{\theta}_o = \Delta\theta_o / \Delta t_o = \Delta\theta_o / (t'_B - t'_A) = \Delta\theta_o / (t'_D - t'_C)$, onde $\Delta t_o = t'_B - t'_A$. Chamaremos $\dot{\theta}_o$ de velocidade angular absoluta, uma vez que $\Delta\theta_o$ e $\Delta t_o$ possuem o mesmo valor em todos os outros referenciais galácticos, excetuando obviamente o referencial *O*.

---

[3] Vale lembrar que as coordenadas 'com linha' se referem às coordenadas medidas no referencial $O'$, enquanto as coordenadas 'sem linha' são medidas no referencial $O$.

Os dois pulsos gravitacionais marcam dois pontos C e D no espaço nos seus eventos finais de chegada nos instantes $t_C$ e $t_D$. Queremos comparar, agora, a medida de deslocamento angular $\Delta\theta_o$ feita em *O'* com a medida de deslocamento angular $\Delta\theta$ feita em *O* entre os mesmos pontos C e D. Vamos supor que o deslocamento angular $\Delta\theta$, entre os pontos C e D, dos referenciais *O'* em relação ao referencial *O* é igual ao deslocamento angular $\Delta\theta_o$ definido pelos mesmos pontos como visto do referencial *O'*. Ou seja, $\Delta\theta = \Delta\theta_o$.

Os intervalos de tempo medidos nos referencias *O* e *O'* para os eventos em C e D são, respectivamente, $\Delta t$ e $\Delta t_o$. Consideremos que os deslocamentos angulares $\Delta\theta$ e $\Delta\theta_o$ ocorrem nestes intervalos como medidos em *O* e *O'*. Para deduzirmos a relação entre eles devemos considerar outros dois eventos. Sejam, agora, dois pulsos gravitacionais localizados na origem de *O'*. O primeiro pulso ocorrendo no instante $t'_o$ e o segundo pulso ocorrendo no instante $t'$. Vamos considerar o intervalo de tempo entre o primeiro e o segundo pulso como sendo $t' - t'_o = \Delta t_o$. E como sabemos, durante este intervalo de tempo houve um deslocamento angular $\Delta\theta_o$ de *O'* em relação a *O*, como medido em *O'*. *Podemos considerar o movimento circular de O', visto em O, como um movimento de translação uniforme numa superfície cilíndrica simples. Dessa forma, é como se o movimento fosse um movimento de translação uniforme sendo realizado num plano (apêndice).* Assim, para a dedução da relação entre $\Delta t$ e $\Delta t_o$ consideraremos as equações de Lorentz que guardam a relação entre as coordenadas de dois referenciais em movimento de translação uniforme no plano. Considerando a relação entre o cilindro e o plano e também as transformações de Lorentz, como estes dois pulsos ocorrem no mesmo ponto do referencial *O'*, então o intervalo de tempo $\Delta t$, medido em *O*, para o deslocamento angular $\Delta\theta_0$, medido em *O'*, deve ser dado por

$$\Delta t = \frac{\Delta t_o}{\sqrt{1 - v^2/c^2}} \quad (2)$$

onde *v* é a velocidade linear de rotação de *O'*, como visto em *O*. Se $\dot\theta$ é a velocidade angular, então $v = \dot\theta r$; onde *r* é a distância entre as origens de *O* e *O'*, medidas em *O*.

Contudo, este deslocamento medido em *O* é $\Delta\theta (= \Delta\theta_o)$. Portanto, a velocidade angular $\dot\theta$ é obtida dividindo o deslocamento angular $\Delta\theta$ do referencial *O'*, como visto em *O*, pelo intervalo de tempo $\Delta t$ em que este deslocamento ocorre.

Dividindo a expressão $\Delta\theta = \Delta\theta_o$ por $\Delta t = \beta \Delta t_o$, após uma pequena álgebra teremos

$$\dot\theta_\pm = \pm \frac{\dot\theta_o}{\sqrt{1 + \frac{\dot\theta_o^2 r^2}{c^2}}} \quad (3)$$

Tomando o sinal positivo[4], vamos propor que esta seja a velocidade de rotação de $O'$ como visto pelo referencial $O$. Então a velocidade linear de rotação das galáxias em relação ao referencial $O$ a uma distância $r$ da origem deste referencial é dada por

$$v = \frac{\dot{\theta}_o r}{\sqrt{1 + \dot{\theta}_o^2 r^2 / c^2}} \tag{4}$$

Esta distribuição de velocidades com relação a distância $r$ ao eixo de rotação obviamente não é válida para corpos rígidos extensos.

## 7. Matéria e Energia Escuras

Em meados do século XX, a exploração do cosmos levou a algumas descobertas de fenômenos que não podiam ser explicados com as teorias vigentes. Em 1929, Hubble catalogou os dados de redshifts de algumas galáxias e descobriu que este desvio para o vermelho aumentava com a distância à nossa galáxia. Os dados usados por Hubble não excluíam, por exemplo, a possibilidade de uma relação quadrática entre o redshift (z) e a distância (R). Dados recentes confirmaram uma relação linear, $cz = H_o R$, entre estas duas grandezas, onde $c$ é a velocidade da luz. O valor do parâmetro de Hubble $H_o$ é estimado atualmente em torno de 73 $km/s/Mpc$[5]. Em 1933, Zwick ao medir as velocidades radiais ($v_r$) de oito galáxias no aglomerado de Coma e estimar a dispersão de velocidades, $\sigma_r = \sqrt{<(v_r - <v_r>)^2>}$, obteve um valor surpreendentemente elevado, porém, próximo aos valores de medidas mais recentes. Valores medidos atualmente estão em torno de 1000 Km/s. Devido a estas descobertas muitas teorias têm sido formuladas, introduzindo conceitos como matéria e energia escura, cuja natureza física permanecem desconhecidas até hoje [8].

Pela teoria exposta aqui afirmamos que estes efeitos podem ser explicados através de um único fenômeno: a rotação dos corpos em relação ao Sistema de Referência em Repouso no Espaço.

A curva de rotação plana das galáxias consiste na distribuição de velocidades tangenciais de seus corpos orbitantes, desde os mais próximos até os mais distantes do seu centro, situados em uma região denominada halo da galáxia. Os corpos situados no halo deveriam obedecer as duas leis de Newton - sua segunda lei de movimento e sua lei de gravitação – de modo que suas velocidades seriam dadas por

$$u = \sqrt{GM/r}, \tag{5}$$

---

[4] A interpretação de $\dot{\theta}_-$, com o sinal negativo, corresponde ao sentido contrário de rotação segundo a convenção adotada.
[5] 1 Mpc (lê-se 'um megaparsec') é uma unidade de medida astronômica e corresponde à distância de aproximadamente $3 \cdot 10^{22}$ metros.

onde *G* é a constante gravitacional de Newton, *M* é a massa da galáxia e *r* é a distância do corpo orbitante ao centro da galáxia. Esta seria a curva esperada com as leis clássicas. Porém, a velocidade angular de rotação em relação ao SR que obtivemos é dada pela expressão (4). Devemos considerar o domínio de pequenas distâncias e/ou velocidades $\dot{\theta}_0 r/c \ll 1$ para o movimento orbital, ou seja, o limite newtoniano. Neste limite identificamos, com o auxílio da equação (5), a velocidade angular $\dot{\theta}_0 = \sqrt{GM/r^3}$. Então, a curva de rotação em relação ao SR esperada deve ser dada pela relação

$$u = \frac{\sqrt{GM/r}}{\sqrt{1+GM/rc^2}}, \qquad (6)$$

que é uma relação que tende a um valor assintótico constante à medida que $GM/rc^2$ vai se tornando muito maior que a unidade.

    O efeito do desvio da luz, quando esta passa próxima a um corpo massivo, pode ser obtido a partir da equação (6). Pois, com esta espressão para a velocidade de um corpo de massa *m* orbitando um outro corpo de massa *M* >>*m* podemos determinar o potencial que produz este perfil de velocidade. Em seguida, encontra-se a métrica a partir deste potencial e, consequentemente, encontra-se também o ângulo de desvio da luz. Esta demonstração pode ser feita posteriormente, sem prejuízo para os fins deste trabalho.

    Por outro lado, o desvio para o vermelho das galáxias distantes tem levado a formulação de algumas hipóteses como "efeito da luz cansada" e principalmente a recessão das galáxias culminando com a teoria do Big-Bang ou versões mais recentes. Através de observações de supernovas do tipo Ia, concluiu-se recentemente que, de fato, quanto maior a distância das galáxias maior é o desvio para o vermelho. Como já relatamos, esta relação é dada por

$$cz = H_o R, \qquad (7)$$

que é uma relação atualmente aceita para as galáxias distantes. Pela teoria exposta aqui, a relação entre os intervalos de tempo entre dois pulsos, medidos em dois referencias dotados de um movimento circular uniforme entre si, é dada pela equação (2). Então a relação entre as freqüências medidas nestes referenciais é dada pelo inverso desta equação. E o redshift *z* associado aos comprimentos de onda dos pulsos deve ser dado por

$$z = \frac{1}{\sqrt{1-v^2/c^2}} - 1$$

    Substituindo a equação (4) na equação anterior, obtemos

$$z = \sqrt{1+\dot{\theta}_o^2 R^2/c^2} - 1$$

Considerando o domínio de grandes distâncias e/ou velocidades $\dot{\theta}_o R/c \gg 1$, então esta relação se torna quase exatamente linear. Com o auxílio da equação (7), podemos identificar $\dot{\theta}_o = H_o$ de forma que teremos

$$z = \sqrt{1 + H_o^2 R^2/c^2} - 1, \qquad (8)$$

que é o redshift das galáxias devido ao seu movimento de rotação - como um todo - em relação ao SR. Observe, agora, que R é a distância do centro da nossa galáxia ao centro da galáxia de onde é proveniente o redshift. Dessa forma, podemos colocar a origem deste SR no centro da Via Láctea. Como vemos, o 'redshift' realmente aumenta com a distância. Para distâncias da ordem de 1 *Mpc*, a relação é aproximadamente quadrática $cz \approx (H_o^2/c)R^2$. Para distâncias bem maiores, a relação vai tendendo assintoticamente para uma relação linear $cz \approx H_o R$.

Esclarecido o que entendemos por matéria e energia escuras, passemos à análise de um efeito que podemos denominar de ovalização das galáxias. Das equações de transformação de Lorentz (1), vemos que se tomamos a medida de comprimento entre dois pontos quaisquer no mesmo instante de tempo no sistema *S*, então a relação entre este comprimento e aquele medido no sistema *S'* é dada por

$$d = \sqrt{1 - v^2/c^2}\, d_o \qquad (9)$$

onde $d_o$ é o comprimento medido em *S'* e *d* é o comprimento medido em *S*.

Introduzindo a identificação $\dot{\theta}_o = H_o$ na relação (4), obtemos a velocidade de rotação das galáxias - como um todo - em relação ao SR, situadas no plano da mesma

$$v = \frac{H_o r}{\sqrt{1 + H_o^2 r^2/c^2}} \qquad (10)$$

Dessa equação, vemos que o movimento das galáxias - como um todo - em relação ao espaço, visto da nossa galáxia, não é nulo. Conseqüentemente, ocorre um achatamento delas por uma razão de $d/d_o$, na direção do seu movimento. Poderíamos denominar este efeito de ovalização das galáxias por torná-las mais ovais. Este efeito, visto por um observador na nossa galáxia e movendo-se junto com ela, dependeria da distância à galáxia na forma

$$\frac{d}{d_o} = \frac{1}{\sqrt{1 + H_o^2 r^2/c^2}},$$

onde ainda utilizamos a equivalência entre o cilindro e o plano para substituir a equação (10) na equação (9).

Para uma galáxia situada no plano da Via Láctea a uma distância de aproximadamente $1 Mpc$ da mesma, o efeito de ovalização seria aproximadamente de

$d/d_o \approx 0,9999$. Para uma distância da ordem de $100\,Mpc$, o efeito seria de $d/d_o \approx 0,9997$. Para uma distância da ordem de $10^4\,Mpc$, teremos $d/d_o \approx 0,3801$. Ou seja, se uma galáxia situada no plano da Via Láctea a uma distância de aproximadamente $10^4\,Mpc$ tem a forma de um disco circular no referencial onde ela está em repouso, daqui da nossa galáxia o diâmetro tangencial é aproximadamente quarenta por cento menor que o diâmetro radial[6].

Por fim, vale ressaltar que para uma relação mais precisa da velocidade dos corpos orbitantes das galáxias, como medidos da nossa galáxia, considere o SR com origem no centro da Via Láctea. Devemos considerar o efeito combinado da velocidade da galáxia - como um todo - em relação a este SR, equação (10), e a velocidade de rotação de seus corpos orbitantes em relação ao centro da galáxia em questão, equação (6). Para isso precisamos de uma lei de composição de velocidades para o movimento circular uniforme, a qual não derivamos aqui.

## 8. Conclusão

A suposição estabelecida aqui de que o sistema de referência em repouso em relação ao observador é um sistema de referência em repouso no espaço precisa ser esclarecida para melhor ser entendida fisicamente. Pois a hipótese leva a crer que existem tantos espaços quanto o número de observadores. Tome, por exemplo, dois observadores em movimento relativo. Todo sistema de referência em repouso em relação a um destes observadores é um sistema de referência em repouso no espaço. Sabemos que IG's em repouso neste sistema detectarão sempre uma velocidade igual a da luz para as ondas gravitacionais. Assim, o observador também está em repouso no espaço. Dessa forma, é como se houvessem vários espaços em superposição coexistindo. A cada observador existiria um espaço no qual ele está em repouso. Esta interpretação seria de fato um absurdo: seria como se houvessem vários éteres coexistindo, se superpondo e em total movimento relativo entre si.

Devemos assumir uma outra forma interpretativa. A nosso ver, existe de fato apenas um único espaço e não vários espaços superpostos. Porém, existe um sistema espaço-temporal associado a cada observador. Neste sentido, existem vários espaços, mas não se superpondo como se poderia pensar. Cada observador não percebe o espaço do outro; o espaço de um observador não coexiste no espaço do outro observador. Embora estes espaços não sejam os mesmos, é evidente que existe uma inter-relação entre eles, uma vez que se um observador realiza medidas de um determinado objeto em seu espaço, então um outro observador no seu próprio espaço realiza medidas também do mesmo objeto.

Para estabelecermos uma comparação entre o Princípio de Relatividade e o Postulado adotado aqui devemos fazer uso principalmente de suas interpretações físicas. Pois interpretações físicas diferentes permitem entender os fenômenos de forma também diferentes. Muitas vezes permitem derivar resultados que não poderiam ser vistos com interpretações anteriores. Como é o caso neste trabalho. Foi possível até mesmo derivar equações para entender, à luz de novas interpretações, as denominadas matéria escura e energia escura. Também foi possível fazer previsões. Um interferômetro de Michelson-

---

[6] Diâmetro tangencial é seu diâmetro na direção do seu movimento circular em relação a nossa galáxia. O diâmetro na direção radial à nossa galáxia é o diâmetro radial.

Morley adaptado às ondas gravitacionais (IG) em movimento relativo a um observador produziria um resultado não nulo para este observador. As galáxias são achatadas na direção do seu movimento em relação a um observador no referencial da nossa galáxia. E as equações de Lorentz também puderam ser derivadas.

    O Postulado que adotamos aqui é mais amplo. Os resultados obtidos com o princípio de relatividade através da teoria da relatividade especial também são válidos aqui. E toda a interpretação decorrente das transformações de Lorentz como dilatação temporal, contração espacial e a noção de simultaneidade. A diferença fundamental entre os dois postulados é que o primeiro introduz a existência de uma infinidade de espaços e tempos. Um espaço e tempo associado a cada observador. Este pode ser considerado o aspecto interpretativo mais relevante da teoria que delineamos aqui. Na teoria de Einstein supõe-se a existência de apenas um espaço e tempo para todos os observadores.

    Este trabalho não pretende ser visto como algo extremamente técnico, no sentido de que seus valores numéricos devam ser esperados à risca, uma vez que outros efeitos podem estar presentes. Vale ressaltar que nos resultados que envolvem valores numéricos, consideramos apenas o movimento no plano da nossa galáxia. As idéias aqui propostas poderão ser desenvolvidas.

## 9. Apêndice

### a. Dilatação do tempo em referenciais com translação uniforme

    Suponha um observador $O'$ que translada em relação a um outro observador $O$ com velocidade constante. E associemos a cada um destes observadores um sistema de coordenadas cartesianas. Podemos agora trabalhar apenas com estes sistemas de coordenadas cartesianas que transladam um em relação ao outro com velocidade constante. Pelo Postulado, adotado aqui, as ondas gravitacionais se propagam em relação a cada referencial com velocidade constante e definida $c$. Como o referencial $O'$ está com velocidade constante $v$ em relação ao referencial $O$, podemos comparar as trajetórias das ondas em cada referencial e equacioná-las como segue.

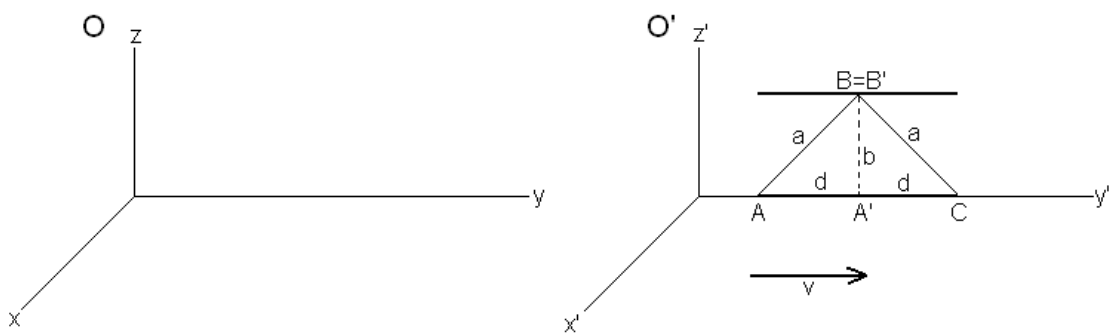

**Figura 1:** Observador $O'$ em movimento de translação uniforme com velocidade *v* em relação a um outro observador *O*.

No referencial $O'$, consideremos um pulso gravitacional que percorre o segmento de reta A'B' reflete em B' e volta pelo mesmo trajeto B'A'≡ A'B' de comprimento *b*. Este segmento pode ser obtido em função da velocidade *c* do pulso e do intervalo de tempo $\Delta t_0$. Este intervalo de tempo $\Delta t_0$ corresponde ao tempo transcorrido no referencial *O'* durante o percurso do pulso no segmento de comprimento *b*. Assim obtemos $b = c\Delta t_0$. No referencial *O*, o pulso é visto percorrer o segmento AB refletir em B e depois prosseguir pelo segmento BC de mesmo comprimento que AB e igual a *a*, com velocidade constante *c*. O tempo transcorrido no referencial *O* durante o percurso do pulso no segmento de comprimento *a* é $\Delta t$. Em termos da velocidade *c* do pulso e do intervalo de tempo $\Delta t$ temos $a = c\Delta t$. Enquanto isto o referencial *O'* transladou o segmento AC de comprimento 2*d* com velocidade constante *v* no intervalo de tempo $2\Delta t$. Assim $d = v\Delta t$. Podemos aplicar o teorema de Pitágoras sobre os comprimentos *a*, *b* e *d* obtendo a relação $a^2 = b^2 + d^2$. De onde encontramos a relação conhecida entre os intervalos de tempo transcorridos nos dois referenciais

$$\Delta t = \frac{\Delta t_0}{\sqrt{1 - v^2/c^2}} \quad ou \quad \Delta t = \beta \Delta t_0$$

Esta é a dilatação temporal encontrada na relatividade especial. Na próxima seção deduzimos esta mesma relação ao considerar um referencial que está em rotação, e com velocidade angular constante, em relação a um outro referencial.

### b. Dilatação do tempo em referenciais com rotação uniforme

Consideremos, agora, o referencial *O'* em rotação uniforme em relação ao referencial *O* com uma velocidade angular constante $\dot\theta$, como vista em *O*. Observamos a trajetória de um pulso gravitacional analogamente ao caso anterior. O pulso, como visto em *O'*, parte do ponto H' com velocidade *c*, percorre a distância *f* do segmento reto H'F' e reflete em F' voltando pelo mesmo caminho até atingir H' novamente. Este percurso é realizado no intervalo de tempo $2\Delta t_0$. Assim podemos fazer $f = c\Delta t_0$. Contudo, no referencial *O* o pulso parte de E, percorre a distância *e* do segmento EF, reflete em F seguindo o segmento FG cujo comprimento é o mesmo de EF e igual a *e*. O intervalo de tempo durante este trajeto é $2\Delta t$. Então pode ser feito $e = c\Delta t$. Enquanto isso, o referencial *O'* percorre a distância 2*g* com velocidade linear $v = \dot\theta r$, como visto em *O*, onde *r* é o raio da órbita circular. Este deslocamento é transcorrido no intervalo de tempo $2\Delta t$. Logo o comprimento *g* é dado por $g = v\Delta t$.

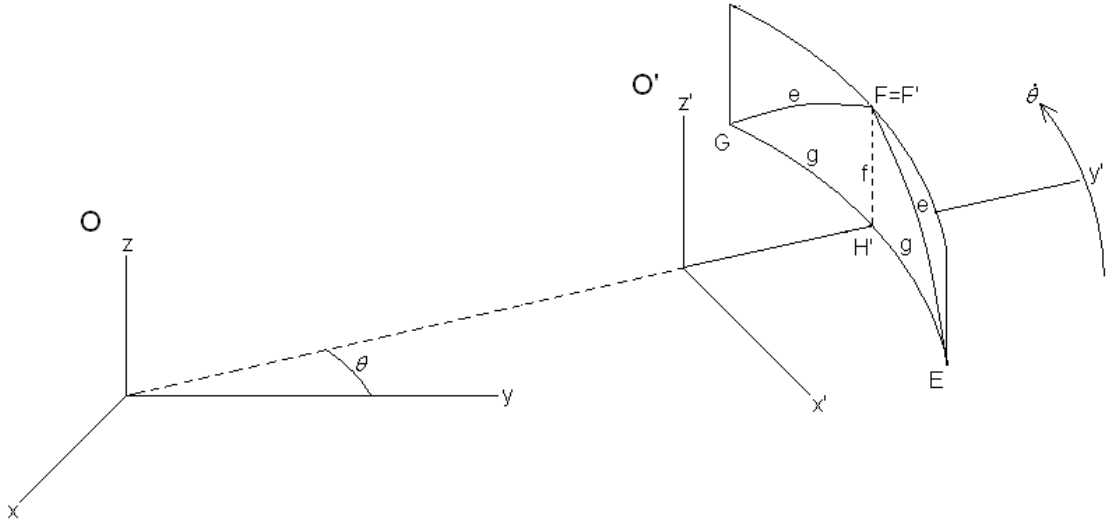

**Figura 2:** Observador $O'$ em movimento de rotação uniforme com velocidade angular $\dot{\theta}$ em relação a um outro observador $O$.

Por isometria [9], podemos transformar os segmentos curvilíneos deste cilindro em retas no plano. Como exemplificado na figura abaixo, seria equivalente a deformar a folha cilíndrica em uma folha plana. Inversamente, esta experiência pode ser elucidada ao desenhar em uma folha de papel plano um triângulo. Em seguida, dobrando-a pode ser vista uma figuração parecida com o desenho à esquerda.

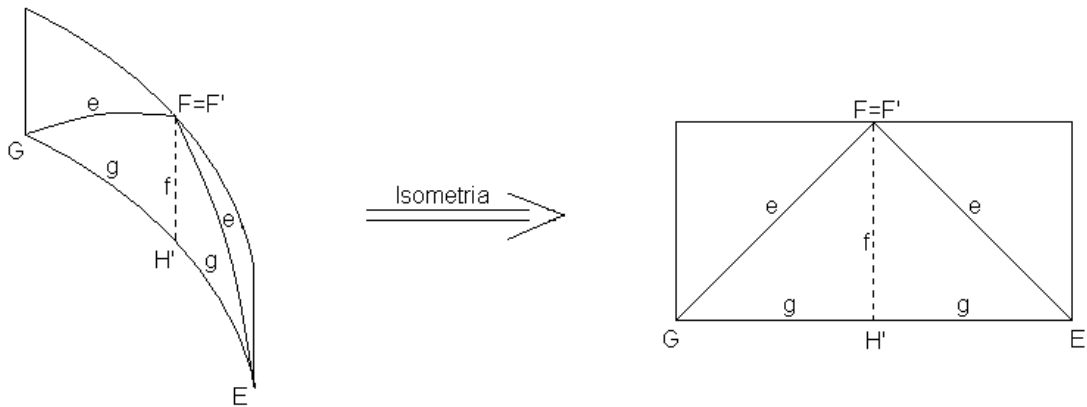

**Figura 3:** Isometria entre o cilindro e o plano. Retas no plano podem ser transformadas em curvas no cilindro e vice-versa.

Usando o teorema de Pitágoras entre os comprimentos $e, f$ e $g$ obtemos a relação $e^2 = f^2 + g^2$. Nestas retas, obtemos a mesma relação temporal que já obtivemos no movimento retilíneo e uniforme. Após uma pequena álgebra vem que

$$\Delta t = \frac{\Delta t_0}{\sqrt{1 - v^2/c^2}} \quad ou \quad \Delta t = \beta \Delta t_0$$

A isometria entre o cilindro e o plano pode guardar ainda mais relações entre o movimento circular e o movimento retilíneo uniforme.